\documentclass[%
 aip,
 apl,
 amsmath,amssymb,
reprint,%
]{revtex4-1}
\usepackage{graphicx}  
\usepackage{dcolumn}   
\usepackage{bm}        
\begin{document}


\title{Mode-filtered electron injection into a waveguide interferometer}

\date{\today}

\author{Sven S. Buchholz}

\author{Ulrich Kunze}
\affiliation{Lehrstuhl f\"ur Werkstoffe und Nanoelektronik, Ruhr-Universit\"at Bochum, 44780 Bochum, Germany}

\author{Saskia F. Fischer}\email[electronic address: ]{Saskia.Fischer@physik.hu-berlin.de}
\affiliation{AG Neue Materialien, Institut f\"ur Physik, Humboldt-Universit\"at zu Berlin, 12489 Berlin, Germany}

\author{Dirk Reuter}

\author{Andreas D. Wieck}
\affiliation{Lehrstuhl f\"ur Angewandte Festk\"orperphysik, Ruhr-Universit\"at Bochum, 44780 Bochum, Germany}

\begin{abstract}
Injection of mode-filtered electrons into a phase-sensitive four-terminal waveguide Aharonov-Bohm (AB) ring is studied. An individually tuneable quantum point contact (QPC) in a waveguide lead of the GaAs/AlGaAs-ring allows to selectively couple to one-dimensional modes in the ring. Thus, we demonstrate single-mode transport in a multi-mode waveguide structure. Coherent mode-filtering by the lowest QPC subband is verified by non-local bend resistance and phase-sensitive AB interference measurements.
\end{abstract}

\maketitle 

Quantum point contacts (QPCs) and electronic waveguides (EWGs) are realizations of a (quasi)-one-dimensional (1D) charge carrier system and show the distinctive property of quantized transverse momentum resulting in conductance quantization.\cite{vanW88,whar88} Therefore, QPCs are applied in fundamental charge and spin transport experiments in mesoscopic physics as electronic beam splitters\cite{nede07}, sensitive single charge\cite{gust08} and spin\cite{folk03} detectors and probes for nonequilibrium dynamics of photogenerated electrons.\cite{hof10} Furthermore, QPCs operated in the ``0.7-conductance-anomaly'' have been discussed to function as (all-electrical) spin polarizers\cite{07anom} in materials of both high\cite{rokh06,debr09} and low\cite{reil02,grah07,yoon07} spin-orbit interaction. It has been suggested that the degree of spin polarization can be probed in a quantum ring - quantum dot device.\cite{hilt10} However, in order to investigate QPCs as spin polarizers, firstly their application as mode filters in the lowest subband calls for experimental realization.

Here, we employed a phase-sensitive waveguide Aharonov-Bohm (AB) ring\cite{buch09a,buch10a,krei10} in order to investigate the mode-filtering properties of QPCs embedded in complex EWG structures. We studied transport in the EWG interferometer in which a QPC is embedded in one of the waveguide leads. The QPC was tuned to the regime of the first and second occupied subbands. By means of bend resistance and electron interference, we show that the selective coupling of (transverse) 1D modes in the EWGs to modes in the adiabatic QPC leads to coherent mode-filtered transport.
\begin{figure}[b]
\begin{center}
\includegraphics[width=1\columnwidth]{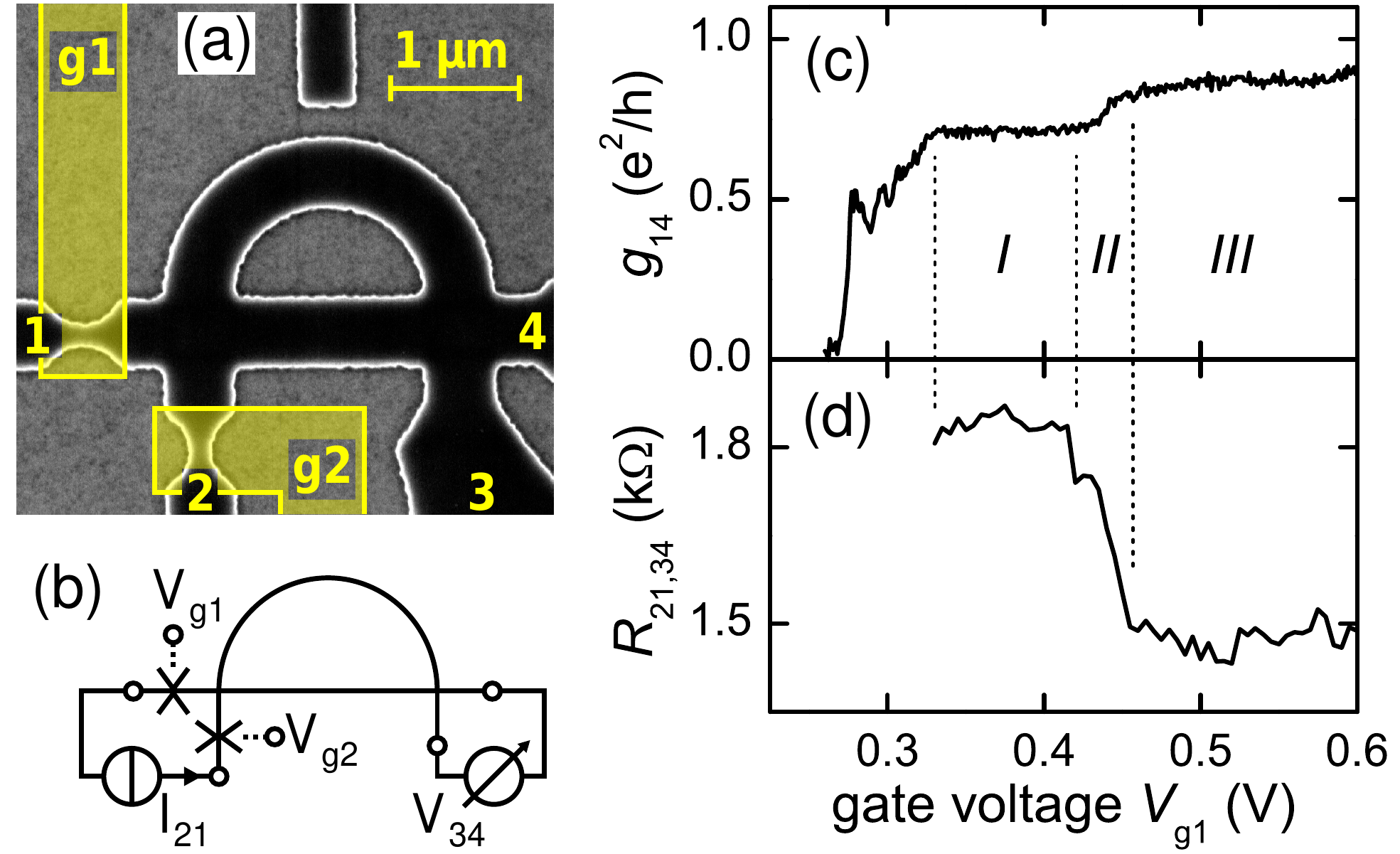}
\caption{(Color online) (a) Scanning electron micrograph of the AB ring with schematically drawn surface gates g1, g2. (b) Non-local measurement configuration ($R_{21,34}$). (c) Two-terminal conductance $g_{14}$ (as measured, no correction for series resistance) across QPC1 and (d) non-local transfer resistance $R_{21,34}$ vs. gate voltage $V_\mathrm{g1}$. The measurement points of (d) were extracted from the raw data of AB measurements at $B=0$ of Fig.~\ref{fig3}. Regions $I$, $II$ and $III$ in (c) and (d) indicate transport in the first ($I$) and second ($III$) QPC subband and the transition in-between ($II$). $T_\mathrm{base}=23$~mK.}
\label{fig1}
\end{center}
\end{figure}

Fig.~\ref{fig1}(a) depicts the asymmetric quantum ring comprised of 520~nm wide EWG arms and leads. In the two left waveguide leads two constrictions (QPCs), $\sim170$~nm wide and $\sim530$~nm long, with metal surface gates g1 and g2 are embedded. From EWG crossing to crossing, the straight and the bent arms of the AB ring are 2.0 and 3.5~$\mu$m long, respectively.

The device was fabricated by electron beam lithography and wet chemical etching from a GaAs/AlGaAs heterostructure with an electron density of $3\times 10^{11}$~cm$^{-2}$ and a mobility of $1\times 10^{6}$~cm$^2/$Vs at $4.2$~K (elastic mean free path $l_\mathrm{e}=9.5$~$\mu$m). Fabrication details can be found elsewhere.\cite{buch09a} 

Transport properties were measured in a single cool-down in a dilution refrigerator at the base temperature $T_\mathrm{base}=23$~mK with standard lock-in technique. Two-terminal characteristics result from current measurements with a voltage excitation of 40~$\mu$V rms. Four-terminal measurements were performed with current excitation of 25~nA at 73.3~Hz and voltage metering in a non-local configuration (Fig.~\ref{fig1}(b)).\cite{buch10a} Magnetotransport was investigated in small ($\left|B\right|<13$~mT) perpendicular magnetic fields with sweep-steps of 30~$\mu$T.

In the EWG arms of the AB ring 3 to 5 transverse 1D modes with a subband separation of $\hbar\omega_{t,\mathrm{EWG}}=$ 0.5 to 2~meV are occupied.\cite{wg} The mode occupation in the QPCs can be tuned by the local gates g1 and g2. While QPC1 was tuned to the first or second subband, QPC2 was kept open at a fixed voltage $V_\mathrm{g2}=650$~mV. 

Fig.~\ref{fig1}(c) shows a two-terminal measurement of the differential conductance $g_{14}$ from terminal 1 to 4 which is typical for a QPC with a high series resistance: $g_{14}$ rises in steps as gate voltage $V_\mathrm{g1}$ increases. The step-like behavior is ascribed to conductance quantization.\cite{vanW88,whar88} Regions $I$ and $III$ are the conductance plateaus of the first and second transverse modes in the QPC, respectively. Region $II$ in-between marks the transition from the first to the second plateau. The reduced plateau height is explained by the high series resistance which originates partly from the alloyed contacts ($\sim10$~k$\Omega$) and partly from the waveguide resistance. 

The discrete subbands of the QPC appear as a signature in the non-local measurement of the transfer (bend) resistance $R_{21,34}$ which is a measure of ballistic transport along the EWG arms of the ring and the waveguide crossings.\cite{buch09a} In Fig.~\ref{fig1}(d) $R_{21,34}$ is roughly constant in regions $I$ and $III$ whereas it drops significantly in the transition region $II$.

In a simple model, a QPC's saddle point potential serves as a 1D mode filter selectively coupling to the modes in the EWGs. Right moving electron waves incident from terminal 1 are coherently transmitted through QPC1 only if the corresponding subband is occupied in the constriction: As an electron wave of transverse mode index $n$ approaches the QPC in transport direction its wavefunction gradually conforms to the constriction conserving the quantum number $n$ and the spin. Thereby, its longitudinal wavenumber $k_\ell$ at the Fermi energy $E_\mathrm{F}$ decreases, whereas the energetic height of the subband minimum as well as the subband separation increase, i.e. kinetic (longitudinal) energy is commuted to potential (offset $E_0$ and transverse) energy:\cite{szaf89,vanH92}
\begin{center}
\vspace{-0.7 cm}
{\small \begin{equation*}
(n-\frac{1}{2})\hbar\omega_{t,\mathrm{EWG}}+\frac{\hbar^2 k_{\ell,\mathrm{EWG}}^2}{2m^*} = E_0+(n-\frac{1}{2})\hbar\omega_{t,\mathrm{QPC}}+\frac{\hbar^2 k_{\ell,\mathrm{QPC}}^2}{2m^*},
\end{equation*}
}
\end{center}
where $\omega_t$ is a measure of the transverse harmonic confinement potential and $\hbar\omega_t$ denotes the quantization energy, i.e. the subband separation. Past the saddle point the electron wave gains $k_\ell$ again and loses potential energy. As a mode of index $n$ it continues to propagate in the EWG ring. Electron waves of modes not occupied in the QPC are reflected as they approach the saddle point. This idealized approach requires an adiabatically varying potential in transport direction to prevent intermode scattering at the orifice, ballistic transport for energy conservation, and a QPC constriction length exceeding the decay length of evanescent modes (subbands not occupied in the QPC).\cite{glaz88,szaf89,vanH92} We assume these requirements to be fulfilled in our device, where the constriction width changes gradually on the scale of the wavelength, the QPC length is larger than its width and the entire ring is shorter than the electrons' mean free path.

The QPC's selective coupling to EWG modes has a significant influence on the transport through the attached AB ring (Fig.~\ref{fig1}(d)). In the non-local measurement configuration, part of the electrons injected through the QPC from terminal 1 travel ballistically along the straight waveguide to terminal 4 (bend resistance).\cite{bara90,buch09a,avis89,kaku91} The probability to reach terminal 4 depends on the transverse mode and the forward-directed longitudinal momentum $\propto k_\ell$. In the lowest mode, electrons have high $k_{\ell,1}$ and their transverse wavefunction maximum is centered in the middle of the waveguide. In the second mode, the two wavefunction maxima are closer to the waveguide boundaries and $k_{\ell,2}<k_{\ell,1}$. 
\begin{figure}[t]
\begin{center}
\includegraphics[width=1\columnwidth]{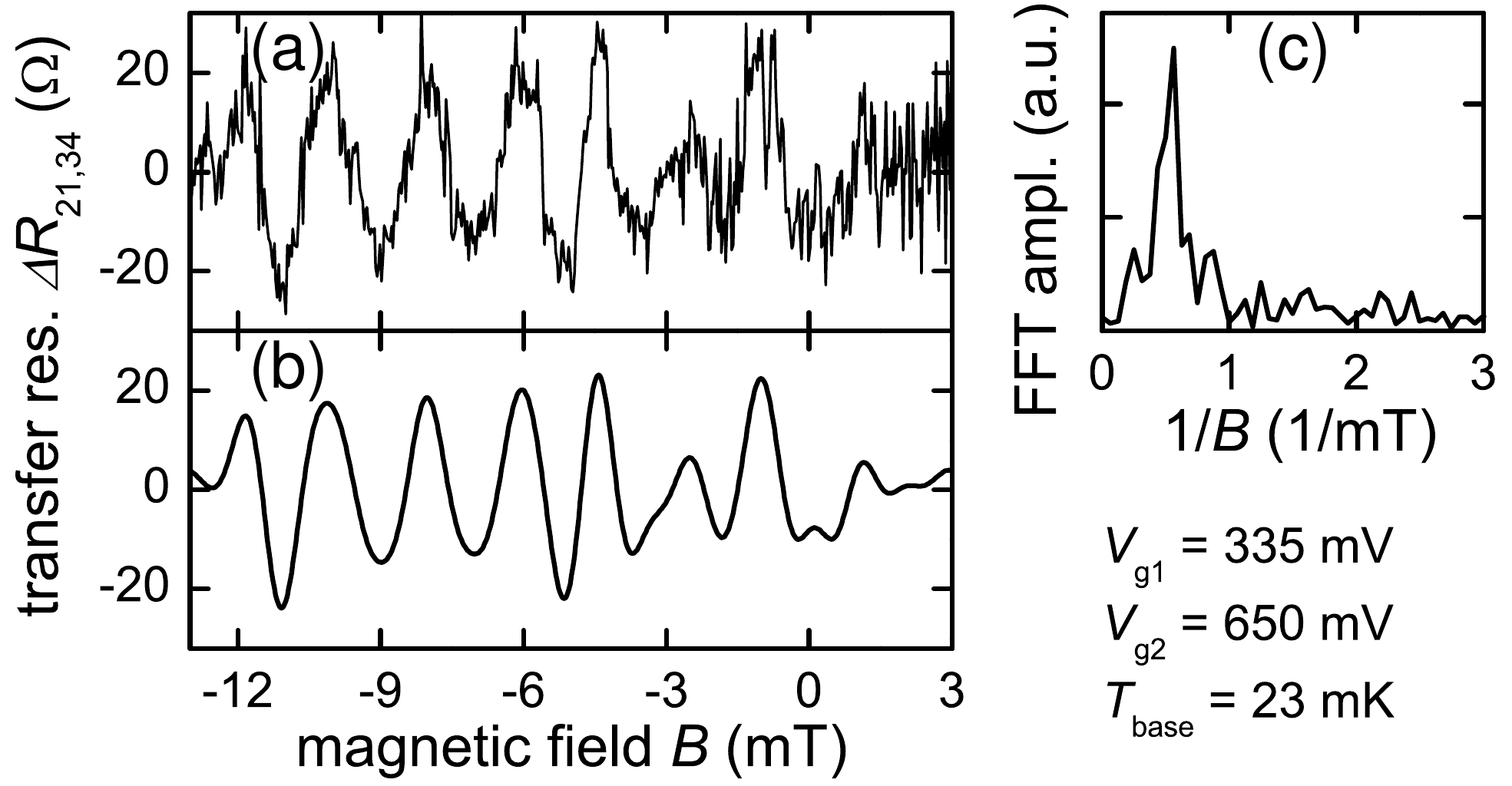}
\caption{Filtering procedure for AB measurements. (a) Raw data of a typical magnetoresistance $\Delta R_{21,34}$ after subtraction of the background (QPC region $I$). (b) Data after FFT-bandpass-filter (0.01 to 1.2~1/mT) as also displayed in Fig.~\ref{fig3}. (c) FFT amplitude of raw data in (a).}
\label{fig2}
\end{center}
\end{figure}
Hence boundary-scattering and scattering in the EWG cross-junctions is increased in the second mode.\cite{krei10,avis89,bara90} The resulting effect is well pronounced at the onset of a new QPC subband: Electrons in the upper subband participate in transport, which leads to an abrupt drop of the total forward transmission along the waveguide crossings. Consequently, $R_{21,34}$ drops abruptly in Fig.~\ref{fig1}(d) with the onset of the second QPC subband (region $II$) as observed in the bend resistance of single EWG cross-junctions.\cite{avis89,bara90,kaku91} Since $k_\ell$ in the EWGs is constant, $R_{21,34}$ does not vary with $V_\mathrm{g1}$ for a fixed number of modes.

The constant $k_\ell$ in the EWGs has an additional consequence. As $V_\mathrm{g1}$ is varied we do not expect a phase shift of the AB interference pattern. A phase-sensitive AB ring detects such an AB resistance oscillation phase shift if $k_\ell$ is varied in one arm of the ring \cite{koba02} or if $k_\ell$ is varied in both arms of an asymmetric ring.\cite{buch10a,krei10}

AB measurements were performed to probe coherence in the different transport regions I and III, i.e. the different QPC injection modes, and to verify the independence of electrons' $k_\ell$ in the EWG on $k_\ell$ in the QPC. A non-local measurement allows for phase-sensitivity in the chosen asymmetric EWG ring design.\cite{buch10a,krei10} Fig.~\ref{fig2} shows a typical magnetotransport measurement of $R_{21,34}$ for $V_\mathrm{g1}=335$~mV (lowest QPC subband, region $I$) and explains our filtering procedure applied to all AB measurements. First, we subtracted the background (gliding average) of each unfiltered measurement (Fig.~\ref{fig2}(a)). The corresponding fast Fourier transform (FFT) is depicted in Fig.~\ref{fig2}(c). The peak at 0.5~1/mT agrees with the expected $h/e$ oscillation period. Second, we removed the high frequency noise superimposed on the $h/e$ and possible $h/(2e)$ oscillations by the application of a FFT bandpass-filter with cutoff frequencies 0.1 and 1.2~1/mT.

\begin{figure}[t]
\begin{center}
\includegraphics[width=1\columnwidth]{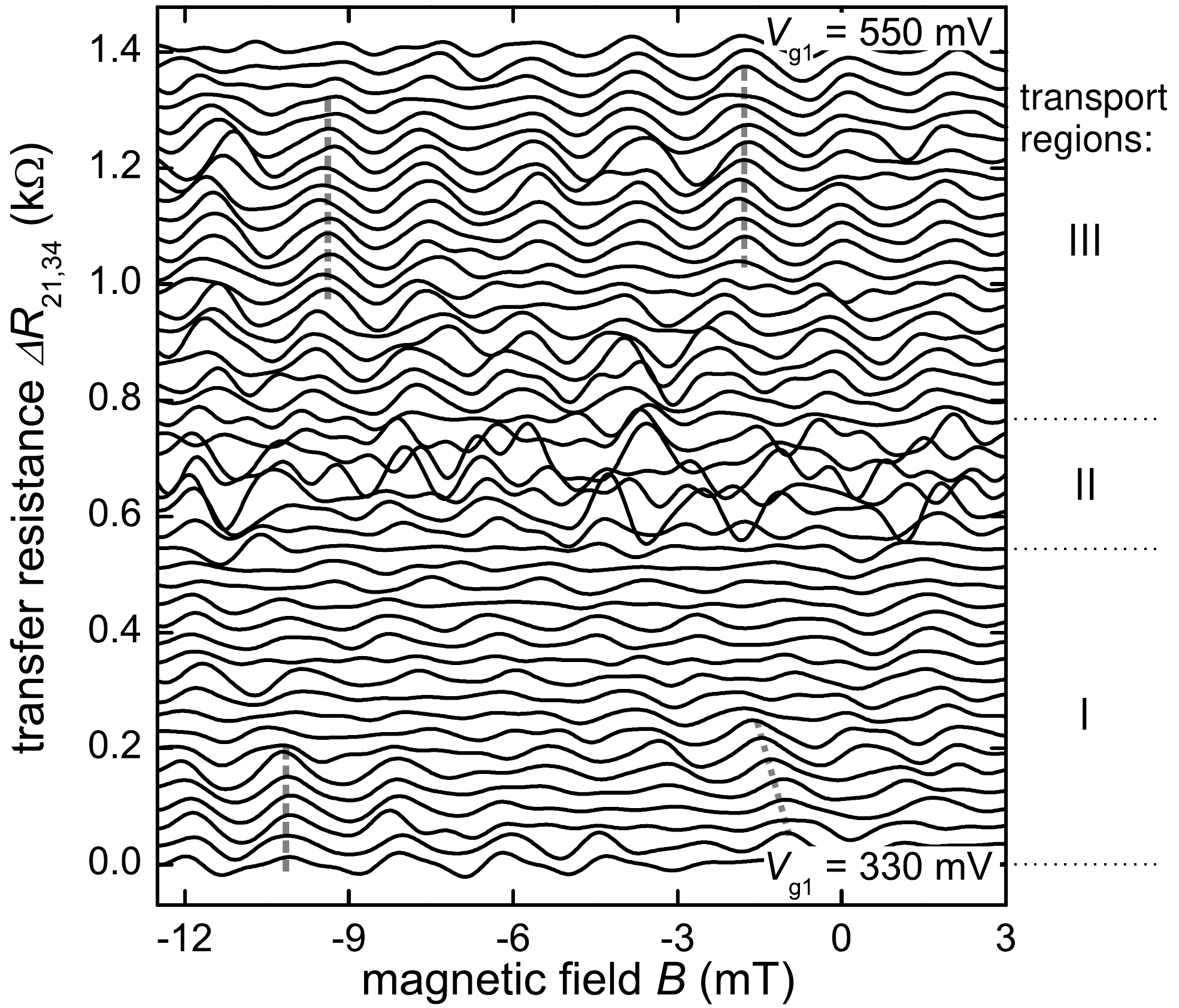}
\caption{AB measurements for QPC transport regions $I$, $II$, $III$ after filtering and offset: $\Delta R_{21,34}$ as a function of magnetic field $B$ for succeeding gate voltages $V_{\mathrm{g1}}$, varied in steps of 5~mV between $V_{\mathrm{g1}}=330$ and 550~mV. $V_{\mathrm{g2}}=650$~mV, $T_\mathrm{base}=23$~mK.}
\label{fig3}
\end{center}
\end{figure}

Further AB measurements of $R_{21,34}$ were recorded for succeeding gate voltages $V_\mathrm{g1}$ from 330 to 550~mV in steps of 5~mV. Fig.~\ref{fig3} shows the filtered data and indicates the three QPC transport regions. For electrons injected via the first ($I$) and the second ($III$) QPC subband the AB oscillations are mostly regular in period (dominating $h/e$ period) and phase relation. However, in region $I$ some deviations are visible: The phase shifts in the lower half of region $I$ for $B>-6$~mT (dotted line), and few curves show $h/(2e)$ periods. Region $III$ shows less deviations from single period oscillations with non-varying phase (dashed lines). The average oscillation amplitude and the visibility\cite{buch10a} in region $III$ are higher than in $I$, namely 24~$\Omega$ and 1.6~$\%$ ($III$) compared to 15~$\Omega$ and 0.8~$\%$ ($I$). Like the bend resistance, the interference amplitude also depends on the transverse electron wavefunction and associated forward and sideward transmission probabilities $T_\mathrm{f}$ and $T_\mathrm{s}$ in the EWG cross-junctions.\cite{krei10} Hence, for electron waves propagating in the lowest EWG subband, high $T_\mathrm{f}$ yields a higher background (bend) resistance (Fig.~\ref{fig1}(d)) but a less pronounced interference amplitude compared to electrons of higher subbands whose $T_\mathrm{s}$ is higher. Consequently, the increased AB resistance amplitude in region $III$ is a further indication for the mode-filtering property of QPC1.

Region $II$ in Fig.~\ref{fig3}, which corresponds to the onset of the second QPC subband, reveals interference patterns with significantly higher AB amplitudes than in regions $I$ and $III$ of which some are distinct $h/(2e)$ oscillations. This irregular behavior suggests a correlation to the onset of the second QPC subband but still remains to be understood in detail. 

In conclusion, we demonstrated coherent mode-filtered electron injection into a waveguide AB ring. A QPC allows for the selective coupling of transport modes in the QPC to modes in the EWG ring, where transport in the lowest mode is of particular interest. Bend resistance and electron interference were measured: While the non-local transfer (bend) resistance drops with the onset of the second QPC subband the corresponding interference amplitude rises, as is expected from considerations of the transverse wavefunction spread and associated transmission probabilities in the waveguide cross-junctions. An AB phase shift is hardly visible indicating a fixed longitudinal momentum $k_\ell$ in the EWG ring.
Hence, proposed investigations with respect to spin and coherence properties\cite{hilt10} appear feasible in the regime of the QPCs' ``0.7-conductance-anomaly''.

The authors appreciate funding by the DFG within the priority program SPP1285. SFF gratefully acknowledges support by the Alexander-von-Humboldt Foundation.


\begin{thebibliography}{aipsamp}
\bibitem{wg} 1D subband occupation and spacing of EWGs was determined from separate measurements on simple waveguides (widths 350 to 550~nm) with and without embedded QPCs.
\end{thebibliography}
\end{document}